\documentclass[prd,aps,eqsecnum,amsmath,floatfix,nofootinbib,preprint,tightenlines]{revtex4}
\usepackage{bm}
\usepackage{epsfig}
\usepackage{float}
\usepackage{color}
\begin{document}
%\preprint{ADP-17-07/T1013, YUPP-I/E-KM-17-02-2}
\thispagestyle{empty}

\title{A ChPT estimate of the strong-isospin-breaking contribution
to the anomalous magnetic moment of the muon}

\author{Christopher L. James}
\email[]{cljames195@gmail.com}
\affiliation{Department of Physics and Astronomy, York University,
4700 Keele St., Toronto, ON CANADA M3J 1P3}
\author{Randy Lewis}
\email[]{randy.lewis@yorku.ca}
\affiliation{Department of Physics and Astronomy, York University,
4700 Keele St., Toronto, ON CANADA M3J 1P3}
\author{Kim Maltman}
\email[]{kmaltman@yorku.ca}
\affiliation{Department of Mathematics and Statistics, York University,
4700 Keele St., Toronto, ON CANADA M3J 1P3}

\altaffiliation{Alternate address: CSSM, Department of Physics,
University of Adelaide, Adelaide, SA 5005 AUSTRALIA}

\begin{abstract}
First-principles lattice determinations of the Standard Model expectation 
for the leading order hadronic vacuum polarization contribution to the 
anomalous magnetic moment of the muon have become sufficiently precise 
that further improvement requires including strong and electromagnetic 
isospin-breaking effects. We provide a continuum estimate of the strong 
isospin-breaking contribution, $a_\mu^{SIB}$, using $SU(3)$ chiral 
perturbation theory. The result is shown to be dominated by resonance-region 
contributions encoded in a single low-energy constant whose value is known 
from flavor-breaking hadronic $\tau$ decay sum rules. Implications of the 
form of the result for lattice determinations of $a_\mu^{SIB}$ are also 
discussed.
\end{abstract}
%\pacs{}
%\keywords{}

\maketitle

\section{\label{intro}Introduction}
The more than $3\ \sigma$ disagreement between the final 2006 BNL E821 
result for $a_\mu$~\cite{Muong-2:2002wip,Muong-2:2004fok,Muong-2:2006rrc}, 
the anomalous magnetic moment of the muon, and subsequent updates of the 
Standard Model (SM) expectation prompted intense interest in improving 
both experimental and theoretical results. Interest in the latter 
has been further heightened by the recently released Fermilab E989 
result~\cite{Muong-2:2021ojo}, which produces an updated experimental world 
average $4.2\, \sigma$ higher than the current best assessment of the SM 
expectation~\cite{Aoyama:2020ynm}.

Hadronic contributions, though representing a small fraction of $a_\mu$,
dominate the uncertainty in the SM prediction. This paper focuses on the
largest of these, the leading-order, hadronic vacuum polarization 
contribution, $a_\mu^{LO,HVP}$. 

As is well known, assuming (as expected) beyond-the-SM contributions 
to experimentally measured $e^+e^-\rightarrow hadrons$ cross sections are
numerically negligible, the SM expectation for $a_\mu^{LO,HVP}$ can be 
obtained as a weighted (``dispersive'') integral over the inclusive 
hadroproduction cross-section ratio $R(s)$. The weight entering this 
integral is exactly known, monotonically decreasing with hadronic 
invariant squared mass, $s$, and strongly emphasizes contributions from 
the low-$s$ region, with $\sim 73\%$ of the full dispersive result coming
from the $\pi\pi$ exclusive mode. Ref.~\cite{Aoyama:2020ynm} provides a 
detailed discussion of the most recent dispersive 
evaluations~\cite{Davier:2019can,Keshavarzi:2018mgv,Keshavarzi:2019abf,Jegerlehner:2017gek}.

A practical complication limiting the accuracy of these determinations is the 
long-standing discrepancy between BaBar~\cite{BaBar:2009wpw,BaBar:2012bdw} 
and KLOE~\cite{KLOE-2:2017fda} $e^+ e^-\rightarrow \pi^+\pi^-$ cross section 
results, which independent determinations by 
CMD2~\cite{CMD-2:2003gqi,CMD-2:2005mvb,CMD-2:2006gxt}, 
BESIII~\cite{BESIII:2015equ}, CLEO-c~\cite{Xiao:2017dqv} and
SND~\cite{SND:2020nwa} have so far failed to resolve. The difference, 
$9.8\times 10^{-10}$~\cite{Keshavarzi:2019abf}, between results for the 
$\pi\pi$ contribution obtained using only BaBar or KLOE in the region 
$0.305\ GeV<E_{CM}< 1.937\ GeV$, and the analogous difference,
$5.5\times 10^{-10}$, between the full $\pi\pi$ contribution obtained using 
averages with either BaBar or KLOE excluded~\cite{Davier:2019can},
both considerably exceed the uncertainty anticipated from the full Fermilab 
E989 experimental program.

The reliance on at-present-discrepant experimental spectral data can, in 
principle, be avoided using lattice results for the electromagnetic (EM) 
current two-point function to evaluate $a_\mu^{LO,HVP}$. This possibility
was first raised in Ref.~\cite{Blum:2002ii} and relies on the alternate 
representation of $a_\mu^{LO,HVP}$ as a weighted integral of the subtracted 
EM vacuum polarization, $\hat{\Pi}_{EM}(Q^2)\equiv \Pi_{EM}(Q^2)-\Pi_{EM}(0)$ 
over spacelike $Q^2\, =\, -s>0$~\cite{Lautrup:1971jf,deRafael:1993za}. While 
the precision of the lattice determination has yet to reach that of the 
dispersive results, there has been rapid progress over the last few
years, with recent updates from the 
BMW~\cite{Budapest-Marseille-Wuppertal:2017okr,Borsanyi:2020mff},
ETMC~\cite{Giusti:2018mdh,Giusti:2019xct,Giusti:2019hkz},
RBC/UKQCD~\cite{RBC:2018dos,Gulpers:2018zha,Lehner:2020crt},
FNAL/HPQCD/MILC~\cite{FermilabLattice:2017wgj,FermilabLattice:2019ugu},
Mainz~\cite{Gerardin:2019rua}, PACS~\cite{Shintani:2019wai} and 
Aubin {\it et al.}~\cite{Aubin:2019usy} collaborations. The most recent 
BMW result~\cite{Borsanyi:2020mff}, in particular, reaches a precision 
of $0.8\%$. While (as detailed, e.g., in Ref.~\cite{Lehner:2020crt}) 
some disagreements persist between results from different lattice 
groups for the dominant $ud$ connected contribution, as well as for the 
$t_0=0.4$ fm, $t_1=1.0$ fm, $\Delta=0.15$ fm RBC/UKQCD ``window'' quantity 
$a_\mu^{ud,conn.,isospin,W}$~\cite{RBC:2018dos}, these disagreements are 
the subject of ongoing scrutiny, and additional sub-$\%$-level lattice 
results are expected in the near future from a number of other lattice groups. 

The current sub-$\%$ precision goal for determining $a_\mu^{LO,HVP}$ on the
lattice necessitates an evaluation of the effects of strong and EM 
isospin-breaking (IB). These receive contributions from both 
quark-line-connected and -disconnected diagrams, with the latter 
much more numerically challenging on the lattice. 

This paper focuses on the strong isospin-breaking (SIB) contribution, 
$a_\mu^{SIB}$. A number of lattice groups have reported determinations 
of the connected contribution, 
$\left[a_\mu^{SIB}\right]_{conn}$~\cite{FermilabLattice:2017wgj,RBC:2018dos,Giusti:2019xct,Lehner:2020crt,Borsanyi:2020mff},
but only one, BMW~\cite{Borsanyi:2020mff}, a result for the disconnected
contribution, $\left[ a_\mu^{SIB}\right]_{disc}$. BMW finds
a strong cancellation between $\left[a_\mu^{SIB}\right]_{conn}$ 
and $\left[ a_\mu^{SIB}\right]_{disc}$, a result anticipated in 
Ref.~\cite{Lehner:2020crt}, which studied the $\pi\pi$ contributions to these
quantities using partially quenched Chiral Perturbation Theory (PQChPT) and
found an exact cancellation of connected and disconnected contributions at
next-to-leading (NLO) chiral order. As we will see below, this cancellation
is specific to NLO, and does not persist to higher order. 
Ref.~\cite{Lehner:2020crt} does not provide a lattice determination of 
$\left[ a_\mu^{SIB}\right]_{disc}$, instead using the NLO PQChPT expression 
for the contribution of the $\pi\pi$ intermediate state as an estimate, 
assigning to this estimate a $50\%$ uncertainty. Results from the literature 
for $\left[a_\mu^{SIB}\right]_{conn}$ and $\left[ a_\mu^{SIB}\right]_{disc}$ 
are summarized in Table~\ref{tab1}. Note that, while (as will be confirmed 
below), one expects finite volume (FV) effects to be small in the full 
connected-plus-disconnected SIB sum, this is not the case for the individual 
connected and disconnected components, and significant FV effects are, 
in fact, observed in the results for $\left[a_\mu^{SIB}\right]_{conn}$ 
reported in Refs.~\cite{RBC:2018dos,Giusti:2019xct,Lehner:2020crt}.

\begin{table}[h]
\begin{center}
\caption{\label{tab1}Lattice results for $\left[a_\mu^{SIB}\right]_{conn}$ 
and $\left[ a_\mu^{SIB}\right]_{disc}$, in units of $10^{-10}$. The $^*$ 
on the disconnected entry from Ref.~\cite{Lehner:2020crt} is a reminder 
that this result is not a lattice one, but rather an estimate of the 
$\pi\pi$ contribution to this quantity obtained using NLO PQChPT, to which
a $50\%$ uncertainty has been assigned. We remind the reader that, while 
FV effects are expected to be small for the connected-plus-disconnected sum,
this is not true of the individual components, and separate connected and 
disconnected results should thus not be compared unless obtained from 
simulations with comparable physical volumes.}

\begin{tabular}{lll}
\hline
$\left[a_\mu^{SIB}\right]_{conn}\times 10^{10}$&\qquad
$\left[ a_\mu^{SIB}\right]_{disc}\times 10^{10}$&\qquad Source\\
\hline
\qquad $9.5(4.5)$&\qquad\qquad --- &\quad\qquad \cite{FermilabLattice:2017wgj,
FermilabLattice:2019ugu}\\
\qquad $10.6(8.0)$&\qquad\qquad --- &\quad\qquad \cite{RBC:2018dos}\\
\qquad $6.0(2.3)$&\qquad\qquad --- &\quad\qquad \cite{Giusti:2019xct}\\
\qquad $9.0(1.4)$&\qquad$-6.9(3.5)^*$&\quad\qquad \cite{Lehner:2020crt}\\
\qquad $6.6(0.8)$&\qquad$-4.7(0.9)$&\quad\qquad \cite{Borsanyi:2020mff}\\
\hline
\end{tabular}
\end{center}
\end{table}

In view of the inflation of the relative error in lattice determinations 
of $a_\mu^{SIB}$ expected from the strong cancellation between connected 
and disconnected contributions, an independent, continuum estimate of this
quantity is of interest. In this paper, we provide such an estimate using
$SU(3)$ chiral perturbation theory (ChPT). 

The rest of the paper is organized as follows. In 
Section~\ref{euclformulation} we set notation, provide the explicit 
expression for $a_\mu^{SIB}$ as a weighted integral over Euclidean $Q^2$ of 
the IB part of the subtracted EM vacuum polarization, $\hat{\Pi}^{SIB}(Q^2)$, 
and discuss the features of this expression which make a ChPT estimate of 
$a_\mu^{SIB}$ feasible. In Section~\ref{chptexpressions}, we provide the 
explicit form of the ChPT representation of $\hat{\Pi}^{SIB}(Q^2)$ needed 
as input to this expression, and outline the flavor-breaking hadronic $\tau$ 
decay sum rule analysis used to determine the input value for a key 
higher-order low-energy constant (LEC) needed to encode the effect of 
$\rho$-$\omega$ mixing. This section also contains our numerical 
results for $a_\mu^{SIB}$. Finally, Section~\ref{conclusions} 
contains a discussion of these results and our conclusions.

\section{\label{euclformulation}The Euclidean integral representation of
$a_\mu^{SIB}$ and feasibility of a ChPT determination}
In what follows, the vector-current two-point functions, $\Pi^{ab}_{\mu\nu}$, 
and associated scalar vacuum polarizations, $\Pi^{ab}$, are defined, as 
usual, by 
\begin{equation}
    \Pi^{ab}_{\mu\nu}(q) = (q_\mu q_\nu - q^2 g_{\mu\nu})\Pi^{ab}(Q^2) =
    i \int d^4 x e^{iq\cdot x} \langle 0 | T \{V^a_\mu (x) V^b_\nu (0)
    \}|0\rangle \, , \label{2point}
\end{equation}
where $Q^2\, \equiv \, -q^2\equiv\, -s$, and $V_\mu^a$ are the members of 
the $SU(3)_F$ octet of vector currents,
\begin{equation}
    V_\mu^a = \bar{q}\frac{\lambda^a}{2}\gamma_\mu q\, .
\end{equation}
The sum of the $u,\, d$ and $s$ contributions to the electromagnetic 
(EM) current then has the standard decomposition,
\begin{equation}
    J_\mu^{EM} = V_\mu^3 + {\frac{1}{\sqrt{3}}}V_\mu^8\, ,
\end{equation}
into $I=1$ ($a=3$) and $I=0$ ($a=8$) contributions, and the vacuum 
polarization, $\Pi_{EM}(Q^2)$, of the two-point function of this 
current the decomposition
\begin{equation}
\Pi_{EM}(Q^2)=\Pi^{33}(Q^2)+{\frac{2}{\sqrt{3}}}\Pi^{38}(Q^2)
+{\frac{1}{3}}\Pi^{88}(Q^2)
\end{equation} 
into pure isovector ($ab=33$), pure isoscalar ($ab=88$), and mixed
isospin ($ab=38$) parts. Since strong isospin-breaking (SIB) is associated
with the $I=1$, $O(m_d-m_u)$ component of the $n_f=3$ QCD mass operator, 
SIB occurs, to leading order in $m_d-m_u$, only in the $38$ part of
$\Pi_{EM}$.

The resulting leading order, $O(m_d-m_u)$ SIB component of the EM current
vacuum polarization is then
\begin{equation}
    \Pi^{SIB}(Q^2) = \frac{2}{\sqrt{3}}\Pi^{38}_{QCD}(Q^2) 
\label{sib_from38} \end{equation}
where the $QCD$ subscript on the right-hand side denotes the 
$O(m_d-m_u)$ QCD contribution and will be dropped in what follows.

\subsection{\label{amusibeuclform}The Euclidean $Q^2$ integral representation
of $a_\mu^{SIB}$}
The full LO, HVP contribution, $a_\mu^{LO,HVP}$, is given, in 
the Euclidean momentum-squared, $Q^2$, representation of 
Refs.~\cite{Lautrup:1971jf,deRafael:1993za}, by the weighted integral
\begin{equation}
a_\mu^{LO,HVP} =\, -4\alpha^2 \int_0^\infty dQ^2 f(Q^2) \hat{\Pi}_{EM}(Q^2)\, ,
\label{amueuclhvp}\end{equation}
with $\hat{\Pi}_{EM}$ the subtracted EM vacuum polarization defined
above, $\alpha$ the EM fine structure constant, and $f(Q^2)$ the 
exactly known kernel
\begin{equation}
f(Q^2) = m_\mu^2 Q^2 Z^3\, {\frac{[1 - Q^2 Z]}{1 + m_\mu^2 Q^2 Z^2}}
\end{equation}
where
\begin{equation}
    Z = {\frac{ \sqrt{Q^4 + 4m_\mu^2 Q^2}-Q^2}{2m_\mu^2 Q^2}}\, .
\end{equation} 
For use in the discussion below, it is convenient to also define the related 
quantity, $a_\mu^{LO,HVP}(Q^2_{max})$, obtained by replacing the upper limit 
of the integral in Eq.~(\ref{amueuclhvp}) by $Q^2_{max}$.

\begin{center}
\begin{figure}[h]
\includegraphics[width=.7\textwidth,angle=0]
{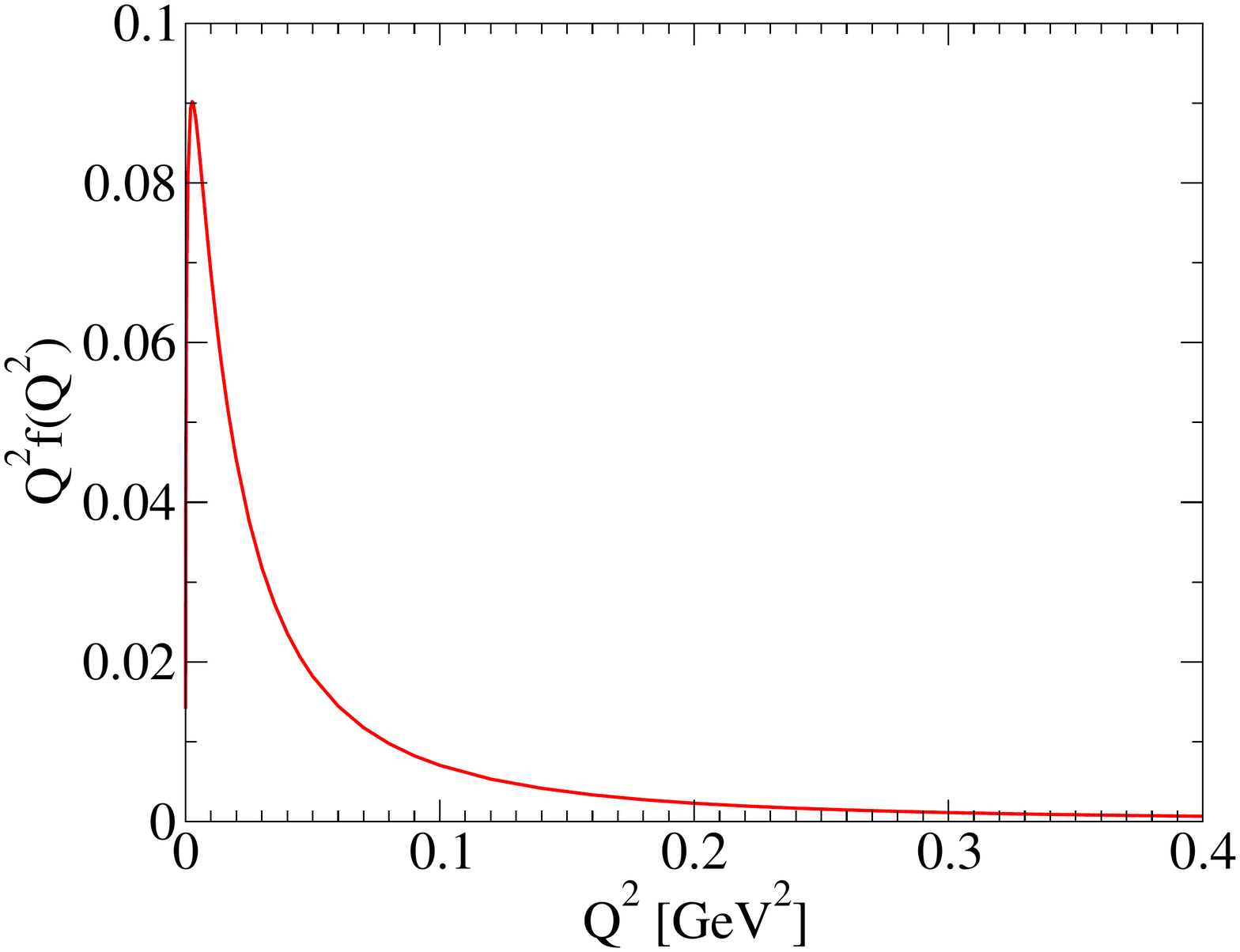}
{\caption{\label{Q2fQ2fig}}The product $Q^2 f(Q^2)$, with $f(Q^2)$ 
the weight appearing in the Euclidean integral representation, 
Eq.~(\ref{amueuclhvp}), of $a_\mu^{LO,HVP}$. }
\end{figure}
\end{center}

The kernel $f(Q^2)$ diverges as $1/\sqrt{Q^2}$ as $Q^2 \rightarrow 0$ and 
falls rapidly with increasing $Q^2$, creating a peak in the integrand of 
Eq.~(\ref{amueuclhvp}) at very low $Q^2\simeq m_\mu^2/4$. At such low
$Q^2$, $\hat{\Pi}_{EM}(Q^2)$ should be very close to linear in $Q^2$, an 
expectation born out by an evaluation of $\hat{\Pi}_{EM}(Q^2)$ using 
$R(s)$ results from Ref.~\cite{Keshavarzi:2018mgv} as input to the 
subtracted dispersive representation
\begin{equation}
\hat{\Pi}_{EM}(Q^2) = \, -\, {\frac{Q^2}{12\pi^2}}\, \int_0^\infty ds\,
{\frac{R(s)}{s(s+Q^2)}}\, .
\label{subdisprep}\end{equation}
The location of the peak of the integrand in Eq.~(\ref{amueuclhvp}) is 
thus essentially just that of the maximum of the product $Q^2 f(Q^2)$. 
Figure~\ref{Q2fQ2fig} shows the behavior of this product as a function
of $Q^2$. Note that an analogous figure for $\hat{\Pi}_{EM}(Q^2) f(Q^2)$,
taking into account the deviation from linearity of $\hat{\Pi}_{EM}(Q^2)$
in the higher-$Q^2$ region, would show an additional suppression, 
increasing with $Q^2$, of contributions at higher $Q^2$ relative to those 
from the region of the peak.

The SIB contribution, $a_\mu^{SIB}$ is, similarly, given, in the
Euclidean-$Q^2$ integral representation, by
\begin{equation}
a_\mu^{SIB} =\, 
-4\alpha^2 \int_0^\infty dQ^2 f(Q^2) \hat{\Pi}^{SIB}(Q^2)\, .
\label{amusibeucl}\end{equation}
As for $\hat{\Pi}_{EM}(Q^2)$, $\hat{\Pi}^{SIB}(Q^2)$ will be very close to 
linear in $Q^2$ in the low-$Q^2$ region, and the maximum of the integrand 
in Eq.~(\ref{amusibeucl}) will thus also occur at $Q^2\simeq m_\mu^2/4$. 

\subsection{\label{feasibility}The feasibility of a ChPT determination}
The fact that the contributions to the integral representation in 
Eq.~(\ref{amusibeucl}) are concentrated at low $Q^2$ raises the 
possibility that a reliable estimate of $a_\mu^{SIB}$ might be 
obtained using the ChPT representation of $\hat{\Pi}^{SIB}(Q^2)$.
An estimate of how reliable such a determination might be can be 
obtained by studying the related $\hat{\Pi}^{33}(Q^2)$ case. 

The utility of this estimate is based on the following similarities 
between the spectral functions, $\rho^{38}(s)$ and $\rho^{33}(s)$,
of $\Pi^{38}(Q^2)$ and $\Pi^{33}(Q^2)$. First, $\rho^{38}(s)$ and 
$\rho^{33}(s)$ share a common threshold, $s=4m_\pi^2$, as well as 
a common saturation of the low-$s$ region by contributions from 
$\pi\pi$ intermediate states. Second, while $\rho^{33}(s)$ is
necessarily $\ge 0$ for all $s$, while $\rho^{38}(s)$ is not, the 
chiral representation of $\rho^{38}(s)$ shows $\rho^{38}(s)$ to be,
like $\rho^{33}(s)$, positive in the low-$s$ $\pi\pi$ region. Third, 
both $\rho^{33}(s)$ and $\rho^{38}(s)$ show sizeable resonance 
enhancements in the $\rho$-$\omega$ region, as evidenced by the 
large $\rho$ peak in the $e^+e^-\rightarrow \pi^+\pi^-$ cross 
sections and the obvious IB interference shoulder, centered at 
$s=m_\omega^2$, on the upper side of that peak. The $\rho$ contribution 
to $\rho^{33}(s)$ is, of course, positive, while the $\rho$-$\omega$
interference contribution to $\rho^{38}(s)$ has a dispersive
shape, with an important contribution which changes sign between
$s<m_\omega^2$ and $s>m_\omega^2$.~{\footnote{See, e.g., Eq. (19),
of Ref.~\cite{Wolfe:2009ts}.} Fits in the interference region 
using various phenomenological models allow one to obtain 
model-dependent separations of the isospin-conserving (IC) $33$ 
and IB $38$ parts of the $\pi\pi$ cross sections. These can be 
converted to the corresponding IC and IB contributions to $R(s)$ 
and the resulting IB contributions integrated with the $a_\mu^{LO,HVP}$ 
dispersive weight to obtain model-dependent estimates of the IB 
$\rho$-$\omega$ interference region contribution to $a_\mu^{LO,HVP}$. 
Such estimates were obtained for a range of models in 
Refs.~\cite{Maltman:2005qq,Wolfe:2009ts,Wolfe:2010gf}. Strong 
cancellations, associated with the change of sign noted above and the 
narrowness of the interference region, were observed in all cases.
These cancellations led to a significantly enhanced model
dependence~\cite{Maltman:2005qq,Wolfe:2009ts,Wolfe:2010gf}. The sign 
of the integrated result was, however, positive over the full range of 
models considered, and hence the same as the sign of the IC $\rho$ 
contribution to $a_\mu^{LO,HVP}$. Integrating, instead, with the 
weight appearing in the subtracted dispersive representation, 
Eq.~(\ref{subdisprep}), one finds, similarly, a common sign for the
IC $\rho$ contribution to $\hat{\Pi}_{EM}(Q^2)$ and the IB $\rho$-$\omega$ 
interference region contribution to $\hat{\Pi}^{SIB}(Q^2)$. From the point 
of view of $\hat{\Pi}^{SIB}(Q^2)$ in the spacelike, $Q^2>0$ region, the
narrow $\rho$-$\omega$ interference contribution to $\rho^{38}(s)$
is essentially indistinguishable from that of a narrow, averaged positive 
contribution located at $s=m_\omega^2$. As far as the subtracted
polarizations are concerned, the spectral functions $\rho^{33}(s)$ 
and $\rho^{38}(s)$ are thus close analogues of one another all the 
way from threshold through the first resonance region, and a study of the 
features of the IC $33$ contribution to the representation 
Eq.~(\ref{amueuclhvp}) can be used to obtain plausible expectations
for the behavior of the corresponding representation, 
Eq.~(\ref{amusibeucl}), of $a_\mu^{SIB}$. 

This observation is of practical use because, in the isospin limit, 
$\hat{\Pi}^{33}(Q^2)={\frac{1}{2}} \hat{\Pi}_{ud;V}(Q^2)$, where 
$\hat{\Pi}_{ud;V}(Q^2)$ is the subtracted polarization of the flavor $ud$, 
$I=1$, vector current, whose spectral function, $\rho_{ud;V}(s)$, has 
been extracted from measured differential non-strange hadronic $\tau$ decay
distributions by ALEPH~\cite{ALEPH:1998rgl,ALEPH:2005qgp,Davier:2013sfa}
and OPAL~\cite{OPAL:1998rrm}. A version of $\hat{\Pi}_{ud;V}(Q^2)$ based 
on the OPAL results for $\rho_{ud;V}(s)$ and the subtracted dispersive 
representation, was constructed in Ref.~\cite{Golterman:2013vca} and 
used to study (i) the convergence of $a_\mu^{LO,HVP}(Q^2_{max})$ to the 
full IC $I=1$ result, $a_\mu^{LO,HVP}$, as $Q^2_{max}$ was increased 
from zero to infinity, and (ii) the utility of various representations 
(including the ChPT representation) of $\hat{\Pi}_{ud;V}(Q^2)$ in the 
low-$Q^2$ region~\cite{Golterman:2013vca,Golterman:2014ksa}. It was 
found that $\sim 82\%$ of the $a_\mu^{LO,HVP}$ arises from 
$Q^2<0.10\ GeV^2$, $\sim 92\%$ from $Q^2<0.2\ GeV^2$, and $\sim 94\%$ 
from $Q^2<0.25\ GeV^2\simeq m_K^2$.{\footnote{See Figures 1 and 2 of 
Ref.~\cite{Golterman:2014ksa} for plots showing the behavior of 
$f(Q^2) \hat{\Pi}_{ud;V}(Q^2)$ as a function of $Q^2$ and 
$a_\mu^{LO,HVP;33}(Q^2_{max})\equiv a_\mu^{33}(Q^2_{max})$ as a 
function of $Q^2_{max}$. Note that the quantity denoted 
$\hat{\Pi}_{ud;V}(Q^2)$ in Ref.~\cite{Golterman:2014ksa} is 
$\Pi_{ud;V}(0)-\Pi_{ud;V}(Q^2)$, and hence differs by an overall 
sign from that used in the current paper.}} With the region between 
$Q^2=0$ and $m_K^2$ plausibly in the range of validity of $SU(3)_F$ 
ChPT, we thus expect that a determination of $a_\mu^{SIB}$ obtained 
using ChPT for $\hat{\Pi}^{SIB}(Q^2)$ and truncating the integral in 
Eq.~(\ref{amusibeucl}) at $Q^2_{max}=0.25\ GeV^2\simeq m_K^2$ will miss 
only $\sim 6\%$ of the total contribution to $a_\mu^{SIB}$, provided 
the ChPT representation used is accurate over this integration region. 

The OPAL-based version of $\hat{\Pi}^{33}(Q^2)$ constructed in 
Ref.~\cite{Golterman:2013vca} can also be used to explore the accuracy
of results obtained using the ChPT representations of subtracted vector
current polarizations in the region up to $Q^2\simeq m_K^2$. To make a 
sensible estimate of the $I=1$ ($33$) contribution to $a_\mu^{LO,HVP}$, 
the chiral order at which the representation of $\hat{\Pi}^{33}(Q^2)$ is 
truncated must be high enough to ensure the effect of the large $\rho$ 
peak in $\rho^{33}(s)$ is incorporated. This contribution first appears in 
the chiral expansion through the next-to-next-to-leading-order (NNLO) LEC, 
$C_{93}$, necessitating the use of the two-loop (NNLO) expression for
$\hat{\Pi}^{33}(Q^2)$. Using this representation, with the value of the
renormalized LEC $C_{93}^r(0.77\ GeV)$ from Ref.\cite{Golterman:2017ljr}} 
as input, one finds an NNLO ChPT estimate for $a_\mu^{33}(0.25\ GeV^2)$ which 
overshoots that produced by the OPAL-based version of $\hat{\Pi}^{33}(Q^2)$
by $\sim 4.8\%$. This slight over-shooting is a consequence of the fact 
that the NNLO representation of $\hat{\Pi}^{33}(Q^2)$ misses small, 
yet-higher-order contributions of the $\rho$ peak to the curvature 
of $\hat{\Pi}^{33}(Q^2)$ in the low-$Q^2$ region. The positivity of 
the $\rho$ contributions to $\rho^{33}(s)$ ensures that these contributions 
would, if included, decrease the magnitude of the resulting representation
of $\hat{\Pi}^{33}(Q^2)$, producing a result for $a_\mu^{33}(0.25\ GeV^2)$ 
lower than that given by the NNLO representation. The (overshooting) 
effect of the truncation at NNLO and the (undershooting) effect of 
omitting contributions from $Q^2>0.25\ GeV^2$ thus work in opposite
directions. The NNLO ChPT estimate, $a_\mu^{33}(0.25\ GeV^2)$, is, in fact,
only $\sim 1.5\%$ below the full ($Q^2_{max}\rightarrow\infty$) $I=1$ 
contribution to $a_\mu^{LO,HVP}$ implied by the OPAL-based dispersive 
version of $\hat{\Pi}^{33}(Q^2)$.

As we will see below, the ChPT result for $a_\mu^{SIB}(0.25\ GeV^2)$ is also
dominated by the contribution of a higher-order LEC encoding resonance-region 
(in this case $\rho$-$\omega$) effects. Since, as noted above, the 
contribution of the $\rho$-$\omega$ interference region to the dispersive
representation of $\hat{\Pi}^{SIB}(Q^2)$ is equivalent to that of a narrow, 
net positive contribution to $\rho^{38}(s)$ located at $s=m_\omega^2$, the 
effect of similarly missing resonance-region-induced, higher-order 
contributions to the low-$Q^2$ curvature of $\hat{\Pi}^{SIB}(Q^2)$ will 
be such that our ChPT estimate for $a_\mu^{SIB}(0.25\ GeV^2)$ will also 
slightly overshoot the true value of this quantity. There will thus, as in the 
case of the NNLO result for $a_\mu^{33}(0.25\ GeV^2)$, be a cancellation 
between the overshooting produced by the use of the truncated ChPT 
representation and the undershooting caused by the truncation of the 
integral representation at $Q^2_{max}=0.25\ GeV^2$. In the analogous
$a_\mu^{33}$ case, these effects are $O(+5\%)$ and $O(-6\%)$, respectively. 
Based on these observations, we expect the combination of the truncation 
in chiral order and truncation of the integral representation at 
$Q^2=0.25\ GeV^2$ to produce an uncertainty of a few to several $\%$ in 
the truncated-in-chiral-order, $a_\mu^{SIB}(0.25\ GeV^2)$ estimate for 
$a_\mu^{SIB}$ obtained below. To be conservative, since this estimate 
for the uncertainty relies on results for the analogous, but not 
identical, $a_\mu^{33}$ case, we assign a significantly expanded 
$10\%$ estimate for the contribution of these effects to the uncertainty 
on the ChPT-based $a_\mu^{SIB}(0.25\ GeV^2)$ estimate for $a_\mu^{SIB}$.

\section{\label{chptexpressions}The ChPT estimate for 
$\hat{\Pi}^{SIB}(Q^2)$}
\subsection{\label{twoloops}$\hat{\Pi}^{SIB}(Q^2)$ to two loops in ChPT}
The forms of the effective $SU(3)_F$ chiral Lagrangian to NLO and 
NNLO were worked out long ago in Refs.~\cite{Gasser:1984gg} and 
\cite{Fearing:1994ga,Bijnens:1999sh}. The two-loop (NNLO) 
representation for the unsubtracted version of the IB polarization, 
$\Pi^{38}(Q^2)$, can be found in Ref.~\cite{Maltman:1995jg}. From this 
expression one finds, recasting the result in terms of the Euclidean 
variable $Q^2= -q^2$, the following result for the subtracted version, 
$\hat{\Pi}^{38}(Q^2)$:
\begin{eqnarray}
&&\hat{\Pi}^{38}(Q^2) = {\frac{\sqrt{3}}{4}}(m_{K^0}^2 - m_{K^+}^2)_{QCD}
\, \bigg[ {\frac{2i\bar{B}(\bar{m}_K^2, Q^2)}{Q^2}}- 
{\frac{1}{48\pi^2 \bar{m}_K^2}} \, +\,\nonumber \\
&&\ \ {\frac{8i\bar{B}(\bar{m}_K^2, Q^2)}{f_\pi^2}}\left(
{\frac{i}{2}}\bar{B}_{21}(m_{\pi}^2, Q^2) + i\bar{B}_{21}(\bar{m}_K^2, Q^2) %
+{\frac{\log \left( m_\pi^2 \bar{m}_K^4/\mu^6\right)}{384\pi^2}}
-L_9^{r}(\mu )\right) \bigg]\, , 
% -2L_9^{r}(\mu )\bigg)  \bigg]\, , 
\label{sub38poln} \end{eqnarray}
where $(m_{K^0}^2 - m_{K^+}^2)_{QCD}$ is the non-EM contribution to the kaon
mass-squared splitting, $\bar{m}_K^2$ is the non-EM part of the average 
physical kaon squared mass, $(m_{K^0}^2+m_{K^+}^2)/2$, $L_9^r$ is the 
usual renormalized NLO LEC of Gasser and Leutwyler~\cite{Gasser:1984gg}, 
$\mu$ is the chiral renormalization scale, $\bar{B}(m^2,Q^2)$ is the standard
subtracted, equal-mass, two-propagator loop function, given, for $Q^2>0$, by
\begin{equation}
    \bar{B}(m^2, Q^2) = \frac{i}{8\pi^2} \bigg[ 1 - \sqrt{1 + 4m^2/Q^2}
\tanh^{-1}\bigg(\frac{1}{\sqrt{1 + 4m^2/Q^2}}\bigg)
    \bigg]\, ,
\end{equation}
and $\bar{B}_{21}$ is the auxillary loop function
\begin{equation}
    \bar{B}_{21}(m^2, Q^2) = \frac{1}{12}\bigg(1 + \frac{4m^2}{Q^2}\bigg)
    \bar{B}(m^2,Q^2) - {\frac{i}{576\pi^2}}\, .
\end{equation}
Our convention for the pion decay constant is that used in 
Ref.~\cite{Gasser:1984gg}, $f_\pi\simeq 92\ MeV$. The first line 
of Eq.~(\ref{sub38poln}) contains the NLO contributions, the second 
line the NNLO contributions. The low-$Q^2$ expansion,
\begin{equation}
{\frac{2i\bar{B}(m^2,Q^2)}{Q^2}}={\frac{1}{48\pi^2 m^2}}\, +\, O(Q^2)\, , 
\end{equation}
has been used in obtaining the subtracted form, Eq.~(\ref{sub38poln}),
from the unsubtracted form given in Ref.~\cite{Maltman:1995jg}. 
The absence of an NLO pion loop contribution in Eq.~(\ref{sub38poln})
reflects the cancellation noted in Ref.~\cite{Lehner:2020crt} 
between NLO $\pi\pi$ intermediate state contributions to the connected and 
disconnected parts of $\hat{\Pi}^{38}$. The presence of the pion loop 
function factor, $\bar{B}_{21}(m_{\pi}^2, Q^2)$, in the NNLO expression 
shows this cancellation does not persist beyond NLO. The result
$L_9^r(\mu =0.77\  GeV)=0.00593(43)$ from Ref.~\cite{Bijnens:2002hp} is
used in obtaining numerical results below.

As is well known, the separation of IB effects into strong and EM 
contributions is ambiguous at $O(\alpha (m_d+m_u))$.{\footnote{A
particularly clear discussion of this point is given in Sections 3.1.1 and 
3.1.2 of the 2019 FLAG report~\cite{FlavourLatticeAveragingGroup:2019iem}.}}
Since $m_d-m_u$ and $m_d+m_u$ differ by only a factor of $\sim 3$ for physical
$m_u$ and $m_d$, this ambiguity is, in fact, at the level of effects second 
order in IB, which we are neglecting. The impact of this ambiguity, in any 
case, lies entirely in the factor $(m_{K^0}^2 - m_{K^+}^2)_{QCD}$ in 
Eq.~(\ref{sub38poln}). At leading order in IB, this factor can be 
determined by subtracting the EM contribution to the K mass-squared 
splitting. This is related to the EM contribution to the pion mass-squared 
splitting by
\begin{equation}
(m_{K^+}^2 - m_{K^0}^2)_{EM} = (m_{\pi^+}^2-m_{\pi^0}^2)_{EM}\, (1+\epsilon_D)
\end{equation}
where $\epsilon_D$ (which depends on the light quark masses) parametrizes 
the breaking of Dashen's Theorem~\cite{Dashen:1969eg}, and is equal to 
zero in the $SU(3)$ chiral limit. Since the experimental pion mass-squared 
splitting receives no SIB contribution at $O(m_d-m_u)$, 
$(m_{\pi^+}^2-m_{\pi^0}^2)_{EM}$ can, up to corrections second order
in IB, be replaced by the corresponding experimental value. Using the
FLAG 2019~\cite{FlavourLatticeAveragingGroup:2019iem} $n_f=2+1+1$ 
result, $\epsilon_D=0.79(7)$, as input, we find
\begin{equation}
(m_{K^0}^2 - m_{K^+}^2)_{QCD}=0.00616(9)\ GeV^2\, ,
\end{equation}
a result valid to first order in IB.

Inputting the NNLO representation of $\hat{\Pi}^{38}(Q^2)$ given by
Eq.~(\ref{sub38poln}) into Eq.~(\ref{amusibeucl}), and using the
numerical input specified above, one finds the following 
results for the NLO and NNLO contributions to $a_\mu^{SIB}(0.25\ GeV^2)$:
\begin{eqnarray}
&&\left[ a_\mu^{SIB}(0.25\ GeV^2)\right]_{NLO}= 0.073\times 10^{-10}
\label{nlosibcont}\\
&&\left[ a_\mu^{SIB}(0.25\ GeV^2)\right]_{NNLO}= 0.552(37)\times 10^{-10}
\label{nnlosibcont}
\end{eqnarray}
where the error on the NNLO contribution 
is that induced by the uncertainty on the input for $L_9^r(0.77\ GeV)$.
The smallness of the NLO contribution in Eq.~(\ref{nlosibcont})
is a reflection of the exact cancellation at NLO between connected and
disconnected contributions from $\pi\pi$ intermediate states. The
total to NNLO, 
\begin{equation}
\left [a_\mu^{SIB}(0.25\ GeV^2)\right]_{NLO+NNLO} = 
0.625(37)\times 10^{-10}\, ,
\end{equation} 
is also small, and dominated by the unsuppressed NNLO contribution. The 
smallness of the NLO+NNLO total should come as no surprise since no LEC 
encoding resonance-region $\rho$-$\omega$ interference contributions to 
$\hat{\Pi}^{38}(Q^2)$ appears in the NNLO representation 
Eq.~(\ref{sub38poln}). The situation is analogous to that of the ChPT
representation of $\hat{\Pi}^{33}(Q^2)$, where the LEC, $C_{93}$,
which encodes the dominant $\rho$ contribution, does not appear in the 
NLO representation. The next subsection addresses this shortcoming of 
the NNLO representation of $\hat{\Pi}^{38}(Q^2)$ and shows how 
results from flavor-breaking hadronic $\tau$ decay sum rules can be
used to quantify the dominant contribution to $a_\mu^{SIB}$ from
terms beyond NNLO in the chiral expansion.

\subsection{\label{beyond2loops}Contributions to $\hat{\Pi}^{SIB}(Q^2)$
beyond two loops}
The mesonic low-energy effective Lagrangian of ChPT has as explicit degrees 
of freedom only the low-lying, pseudoscalar mesons. The effects of 
resonance degrees of freedom, which have been integrated out, are encoded 
in the LECs of the effective theory. As is well known, contributions from 
the lowest-lying resonances provide estimates for these LECs which 
typically agree well with phenomenological determinations~\cite{Ecker:1988te}. 

At low $Q^2$, the $\rho$-$\omega$ mixing 
contribution to $\rho^{38}$ produces a leading low-$Q^2$ contribution 
to $\hat{\Pi}^{38}(Q^2)$ of the form $C_{\rho \omega} Q^2$ where 
$C_{\rho \omega}$ is a constant proportional to the product 
$f_\rho f_\omega \theta_{\rho \omega}$, with $f_\rho$ the $\rho$ 
decay constant (which parametrizes the $\rho$ coupling to $J_\mu^3$), 
$f_\omega$ the $\omega$ decay constant (which parametrizes the $\omega$ 
coupling to $J_\mu^8$) and $\theta_{\rho \omega}$ the IB parameter
characterizing the strength of $\rho$-$\omega$ mixing. No tree-level
contribution of the form $C Q^2$ appears in the NNLO expression
Eq.~(\ref{sub38poln}), establishing that $\rho$-$\omega$ mixing effects
are not yet encoded in the NNLO form. The reason for this absence is
obvious. An operator in the effective Lagrangian producing an SIB, 
tree-level $C Q^2$ contribution to $\hat{\Pi}^{38}(Q^2)$ would have 
to include one factor of the quark mass matrix and four derivatives 
(two to produce the factor $(q_\mu q_\nu-g_{\mu\nu}q^2)$ in 
$\Pi^{38}_{\mu\nu}$ and two to produce the $Q^2$ in the $CQ^2$ contribution
to $\hat{\Pi}^{38}(Q^2)$). Such an operator is NNNLO in the chiral counting. 
The LECs encoding the effects of $\rho$-$\omega$ mixing (as well as of 
all other higher-energy degrees of freedom integrated out in forming the 
effective Lagrangian) thus do not appear in the chiral expansion 
of $\hat{\Pi}^{38}(Q^2)$ until NNNLO. 

Model-dependent results for the contribution to $a_\mu^{SIB}$ from the 
$\rho$-$\omega$ interference region can, of course, be obtained using 
experimental results for the $\pi\pi$ cross-sections in the interference
region and separations of the IC and IB contributions to these cross sections
produced by fits based on phenomenological models of the pion form 
factor $F_\pi (s)$. Such results, of course, provide no information about 
NNNLO (and higher) contributions to $a_\mu^{SIB}$ from other high-energy 
degrees of freedom also integrated out in forming the effective Lagrangian, 
though they do serve to provide an estimate of the expected scale of 
NNNLO and higher order contributions. Fits to a range of experimental 
$\pi\pi$ cross-section data sets involving models for which the resulting 
$\chi^2/dof$ was $<1$ were found to produce $\rho$-$\omega$ mixing 
contributions between $\sim 2\times 10^{-10}$ and 
$\sim 5\times 10^{-10}$~\cite{Wolfe:2009ts,Davier:2010fmf}, confirming
the numerical importance of contributions beyond NNLO. Contributions 
other than that induced by $\rho$-$\omega$ mixing, for example due to
$\rho^\prime$-$\omega^\prime$ mixing, are, of course, also expected at 
some level. With the $\rho^\prime$ and $\omega^\prime$ having comparable 
widths, and no analogue of the $\rho$-$\omega$ interference shoulder 
evident in the $\pi\pi$ cross-sections in the $\rho^\prime$-$\omega^\prime$ 
region, no similar phenomenological estimate is possible for such higher 
resonance contributions. 

An advantage of the chiral representation of the low-$Q^2$ contributions
to $a_\mu^{SIB}$ is that contributions from all degrees of freedom 
integrated out in forming the effective Lagrangian, not just those from
the $\rho$-$\omega$ interference region, will be encoded in the relevant 
NNNLO (and higher) LECs. It turns out that, at NNNLO, there is only one 
such LEC, denoted $\delta C_{93}^{(1)}$ in Ref.~\cite{Golterman:2017ljr}. 
Retaining only vector external sources, $v_\mu = v_\mu^a \lambda^a/2$, 
the form of the associated NNNLO term in the effective Lagrangian needed 
to generate tree-level contributions to vector-current two-point functions 
reduces to 
\begin{equation}
8B_0 Q^2 \delta C_{93}^{(1)}\, Tr\left[M v^\mu v^\nu \right]\,
\left( q_\mu q_\nu - g_{\mu\nu}q^2\right)\, ,
\label{nnnlolecoperatorform}\end{equation}
where $M$ is the quark mass matrix and $B_0$ the standard leading-order 
(LO) LEC, related to the chiral limit value of the quark condensate. 
The tree-level contribution to $\Pi^{38}_{\mu\nu}$ and thence to 
$\hat{\Pi}^{38}$ is obtained by taking the second derivative of this 
expression with respect to $v_\mu^3$ and $v_\nu^8$. An estimate of 
beyond-NNLO contributions to $a_\mu^{SIB}$ thus requires only a 
determination of the LEC $\delta C_{93}^{(1)}$. 

The situation for the chiral representation of $a_\mu^{SIB}$ is similar
to that of the chiral representation of the $I=1$ ($ab=33$) contribution 
to $a_\mu^{LO,HVP}$, where the leading (tree-level LEC) contribution from the 
$\rho$ resonance enters beginning only at NNLO. The NLO representation thus
produces a dramatic underestimate of $a_\mu^{LO,HVP;33}$. As noted above, 
this underestimate is almost completely cured once NNLO contributions, 
including, in particular, the $\rho$-dominated contribution proportional 
to $C_{93}$, are included.

It turns out that the NNNLO LEC, $\delta C_{93}^{(1)}$, which encodes the 
contributions to $a_\mu^{SIB}$, at NNNLO, from all degrees of freedom 
integrated out in forming the effective Lagrangian (including those from the 
$\rho$-$\omega$ interference region) has already been determined in a 
flavor-breaking (FB), inverse-moment finite-energy sum rule (IMFESR) 
analysis of non-strange and strange hadronic $\tau$ decay distribution 
data~\cite{Golterman:2017ljr}. We outline this determination below, and 
provide a numerical update of its results for $\delta C_{93}^{(1)}$.

FB hadronic $\tau$ data can be used to determine $\delta C_{93}^{(1)}$ 
because of the close relation between $\hat{\Pi}^{38}(Q^2)$ and the FB 
vector current combination $\hat{\Pi}_{ud-us;V}(Q^2)\equiv 
\hat{\Pi}_{ud;V}(Q^2)\, -\, \hat{\Pi}_{us;V}(Q^2)$.{\footnote{Since 
$m_s\not= m_u$, the flavor $us$ vector current is not conserved. The 
associated two-point function thus has non-zero spin $J=1$ and $0$ 
vaccum polarizations, each of which has a kinematic singularity at 
$Q^2=0$. As usual, these singularities cancel in the $J=0+1$ sum, and 
by $\hat{\Pi}_{us;V}(Q^2)$ we mean the subtracted version of the 
kinematic-singularity-free sum of the $J=0$ and $1$ polarizations.}} 
$\hat{\Pi}_{ud-us;V} =\hat{\Pi}^{11}+\hat{\Pi}^{22}-\hat{\Pi}^{44}
-\hat{\Pi}^{55}$, and hence involves symmetric products of flavor-octet 
vector currents. The FB component of the QCD quark mass operator
\begin{equation}
{\frac{-2}{\sqrt{3}}} (m_s-m_u-m_d)\, \bar{q}{\frac{\lambda^8}{2}}q
\end{equation}
is proportional to the $a=8$ member of the flavor octet, 
$S^a=\bar{q}{\frac{\lambda^2}{2}}q$, of light-quark scalar densities.
The FB combination $\hat{\Pi}_{ud-us;V}$ thus, to first order in FB,
is determined by the $a=8$ member of the symmetric $8_F$ multiplet
of the products of octet vector currents. Since the SIB component 
of the QCD quark mass operator
\begin{equation}
-(m_d-m_u)\, \bar{q}{\frac{\lambda^3}{2}}q
\end{equation}
is proportional to the $a=3$ member of the same octet of scalar
densities, and $\Pi^{38}_{\mu\nu}$ involves the symmetric product, 
$J_\mu^3 J_\nu^8 +J_\mu^8 J_\nu^3$, of two members of the same octet
of vector currents, $\hat{\Pi}^{38}$, is determined, to first order in SIB,
by the $a=3$ member of the same symmetric $8_F$ multiplet of products of 
the octet vector currents. A determination of the contributions beyond NNLO 
to $\hat{\Pi}_{ud-us;V}$ will thus, up to corrections higher order in 
$SU(3)_F$ breaking, also provide a determination of the contributions 
beyond NNLO to $\hat{\Pi}^{38}$.

The NNNLO version of the relation between these two quantities follows
immediately from the structure of the NNNLO operator in 
(\ref{nnnlolecoperatorform}). The FB NNNLO contribution to 
$\hat{\Pi}_{ud-us;V}(Q^2)$ and SIB NNNLO contribution to 
$\hat{\Pi}^{SIB}(Q^2)$ produced by this operator are
\begin{equation}
\left[ \hat{\Pi}_{ud-us;V}(Q^2)\right]_{NNNLO,LEC}
= -\, 8 Q^2(m_K^2-m_\pi^2)\, \delta C_{93}^{(1)}
\label{fbnnnlopihat}\end{equation}
and
\begin{equation}
\left[ \hat{\Pi}^{SIB}(Q^2)\right]_{NNNLO,LEC}
= -\, {\frac{8}{3}} Q^2 \left( m_{K^0}^2 - m_{K^+}^2\right)_{QCD}
\, \delta C_{93}^{(1)}
\label{sibnnnlopihat}\end{equation}
where the LO relations $B_0(m_s-m_u)=m_K^2-m_\pi^2$ and 
$B_0(m_d-m_u)=\left( m_{K^0}^2 - m_{K^+}^2\right)_{QCD}$ have
been used to recast the results in terms of pseudoscalar meson
masses. While (since they encode resonance-region contributions missing 
at NNLO) we expect these terms to dominate the contributions beyond NNLO, 
the argument above shows that the relation between NNNLO and higher FB 
contributions to $\hat{\Pi}_{ud-us;V}(Q^2)$ and NNNLO and higher SIB 
contributions to $\hat{\Pi}^{SIB}(Q^2)$ is more general, and extends 
beyond the relation between the tree-level NNNLO contributions. 

We now outline the determination of $\delta C_{93}^{(1)}$ from the FB 
IMFESR analysis of hadronic $\tau$ decay data. This analysis is favored
as a means of determining $\delta C_{93}^{(1)}$ because the spectral 
functions, $\rho_{ud;V}(s)$ and $\rho_{us;V}(s)$, of $\hat{\Pi}_{ud;V}$ 
and $\hat{\Pi}_{us;V}$ can be determined experimentally, up to $s=m_\tau^2$, 
from the measured differential non-strange and strange hadronic $\tau$ decay 
distributions~\cite{Tsai:1971vv}. Experimental data can thus be used 
to evaluate the first term on the right-hand side of the FB IMFESR 
\begin{eqnarray}
\label{IMFESR}
{\frac{d\hat{\Pi}_{ud-us;V}(Q^2)}{dQ^2}}\Big|_{Q^2=0}\, =\,
&=& -\, \int_{4m_\pi^2}^{s_0} ds\, w_\tau (s/s_0)\, {\frac{\rho_{ud;V}(s)
-\rho_{us;V}(s)}{s^2}}\\
&&-{\frac{1}{2\pi i}}\oint_{|s|=s_0} ds\, w_\tau (s/s_0)\,
{\frac{\hat{\Pi}_{ud-us;V}(Q^2=\, -s)}{s^2}}
\, ,
\label{fbimfesrslope}
\end{eqnarray}
provided $s_0\leq m_\tau^2$. The operator product expansion (OPE) is used
to evaluate the (numerically very small) second term on the right-hand side. 
The $\tau$ kinematic weight factor, $w_\tau (x)=1-3x^2+2x^3$, with $x=s/s_0$, 
has been included (i) because of its double zero at $s=s_0$, which serves to 
suppress duality violating contributions and improve the accuracy of the OPE 
approximation~\cite{Maltman:1998uzw,Dominguez:1998wy}, and (ii) because 
its derivative with respect to $s$ at $s=0$ is $0$, which ensures 
only the derivative of the polarization with respect to $Q^2$ appears on 
the left-hand side. Analogous IMFESRs provide the slopes with respect to 
$Q^2$, at $Q^2=0$, of the separate non-strange and strange polarizations 
$\hat{\Pi}_{ud;V}$ and $\hat{\Pi}_{us;V}$. The chiral representations of 
$\hat{\Pi}_{ud;V}(Q^2)$ and $\hat{\Pi}_{us;V}(Q^2)$ are known to NNLO
and given in Ref.~\cite{Amoros:1999dp}. Both contain numerically small 
NLO and NNLO loop contributions and a common, numerically dominant 
tree-level NNLO LEC contribution $8Q^2 C_{93}^r$ encoding the leading 
$\rho$ contribution to $\hat{\Pi}_{ud;V}(Q^2)$ and $K^*$ contribution 
to $\hat{\Pi}_{us;V}(Q^2)$. These leading representations of 
resonance-region effects cancel in the NNLO representation of the FB 
difference $\hat{\Pi}_{ud-us;V}(Q^2)$. Resonance-region contributions 
to $\hat{\Pi}_{ud-us;V}(Q^2)$ thus, as for $\hat{\Pi}^{38}(Q^2)$ (and 
for the same reason as in the $\hat{\Pi}^{38}(Q^2)$ case) first enter 
at NNNLO in the chiral expansion. Contributions to the slopes with respect 
to $Q^2$ of $\hat{\Pi}_{ud;V}(Q^2)$ and $\hat{\Pi}_{us;V}(Q^2)$ in the 
low-$Q^2$ region are expected to be dominated by the effects of the 
$\rho$ and $K^*$ resonances. Since these contributions produce slopes 
at $Q^2=0$ which, in the narrow width approximation, are proportional 
to $f_\rho^2/m_\rho^4$ and $f_{K^*}^2/m_{K^*}^4$, a FB difference of order 
$\sim 40\%$ between the $\hat{\Pi}_{ud;V}(Q^2)$ and $\hat{\Pi}_{us;V}(Q^2)$ 
slopes would not be unexpected. A difference of this magnitude is easily 
determinable from the FB IMFESR, Eq.(\ref{fbimfesrslope}), given the 
accuracy of current experimental hadronic $\tau$ decay distributions. 

The slope ${\frac{d\hat{\Pi}_{ud-us;V}(Q^2)}{dQ^2}}\Big|_{Q^2=0}$, was 
determined in Ref.~\cite{Golterman:2017ljr} using then-current OPE input 
and $\rho_{ud;V}(s)$ and $\rho_{us;V}(s)$ obtained from then-current 
versions of the non-strange and strange experimental $\tau$ decay 
distributions. Important inputs to this analysis are the exclusive-mode 
strange $\tau$ branching fractions (BFs), which set the overall scales 
of the corresponding exclusive-mode contributions to $\rho_{us;V}(s)$. 
At the time of the analysis of Ref.~\cite{Golterman:2017ljr}, there 
was a disagreement between the HFAG assessments of the two 
$\tau\rightarrow K\pi\nu_\tau$ BFs and the expectations for these BFs 
from the dispersive analysis of Ref.~\cite{Antonelli:2013usa} (ACLP). 
Since the sum of these BFs sets the normalization for the dominant $K\pi$ 
contribution to $\rho_{us;V}(s)$, this disagreement produced a disagreement 
between results for the FB slope at $Q^2=0$ obtained using the HFAG and 
ACLP $K\pi$ normalizations. Ref.~\cite{Golterman:2017ljr} thus quoted two 
different determinations of the FB slope difference, and hence two different 
results for $\delta C_{93}^{(1)}$, the latter obtained assuming the 
slope difference is dominated by the NNNLO contribution. 

New experimental information has since resolved the $K\pi$ BF discrepancy 
in favor of the dispersive ACLP expectation: the sum of the 
$\tau\rightarrow K\pi\nu_\tau$ BFs reported in the 2019 HFLAV 
compilation~\cite{HFLAV:2019otj} agrees well with the ACLP expectation and,
in addition, has a significantly smaller uncertainty. We have thus updated
the determination of $\delta C_{93}^{(1)}$ in Ref.~\cite{Golterman:2017ljr} 
using (i) current 2019 HFLAV results for all $\tau$ BFs and correlations, 
(ii) the updated determination of $\rho_{ud;V}(s)$ reported in 
Ref.~\cite{Boito:2020xli}, (iii) updated PDG~\cite{ParticleDataGroup:2020ssz} 
input for $\alpha_s$, $V_{ud}$ and $V_{us}$, (iv) updated 2019 
FLAG~\cite{FlavourLatticeAveragingGroup:2019iem} input for the light-quark 
masses, and (v) the most recent HPQCD result~\cite{Davies:2018hmw} for the 
strange-to-light-quark condensate ratio. While included for completeness, 
updates other than those to the $\tau\rightarrow K\pi\nu_\tau$ BFs have 
negligible impact on the results for the FB slope difference. The updated 
result
\begin{equation}
{\frac{d\hat{\Pi}_{ud-us;V}(Q^2)}{dQ^2}}\Big|_{Q^2=0}\, =\,
-0.0862(24)\ GeV^{-2}
\end{equation}
has an improved error and central value very close to the ACLP-based 
result, $-0.0868(40)\ GeV^{-2}$, of Ref.~\cite{Golterman:2017ljr}. The
updated slope produces an updated estimate
\begin{equation}
\delta C_{93}^{(1)}\, \left(m_K^2-m_\pi^2\right)\, =\, 0.00534(37)
\ GeV^{-2}
\label{updateddelc931}\end{equation}
for the NNNLO LEC $\delta C_{93}^{(1)}$.

Our assessment of the NNNLO contribution to $\hat{\Pi}^{SIB}(Q^2)$ is
obtained by substituting Eq.~(\ref{updateddelc931}) into 
(\ref{sibnnnlopihat}). Weighting this expression with the factor
$-4\alpha^2 f(Q^2)$ appearing in Eq.~(\ref{amusibeucl}) and integrating
between $Q^2=0$ to $0.25\ GeV^2$ produces our estimate,
\begin{equation}
\left[ a_\mu^{SIB}(0.25\ GeV^2)\right]_{NNNLO} =2.69(18)\times 10^{-10}\, ,
\label{nnnlolecamusibresult}\end{equation}
for the NNNLO contribution to $a_\mu^{SIB}(0.25\ GeV^2)$, and hence for
the NNNLO contribution to $a_\mu^{SIB}$. The error in
Eq.~(\ref{nnnlolecamusibresult}) reflects only the uncertainty on
the input for $\delta C_{93}^{(1)}$ from Eq.~(\ref{updateddelc931}).
We assign an additional $\sim 30\%$ uncertainty to the NNNLO result
to account for the absence of small non-resonance-induced NNNLO loop 
contributions and the impact of possible contributions higher order in FB 
to the slope at $Q^2=0$ of $\hat{\Pi}_{ud-us;V}(Q^2)${\footnote{The
FB IMFESR provides an essentially purely experimental determination of the 
FB slope difference. The associated determination of $\delta C_{93}^{(1)}$, 
however, relies on the assumption that this result is dominated by the 
leading-order-in-FB contribution associated with the NNNLO operator
(\ref{nnnlolecoperatorform}). This assumption might be subject to
$O(30\%)$ $SU(3)_F$ corrections.}} 

\begin{center}
\begin{figure}[h]
\includegraphics[width=.8\textwidth,angle=0]
{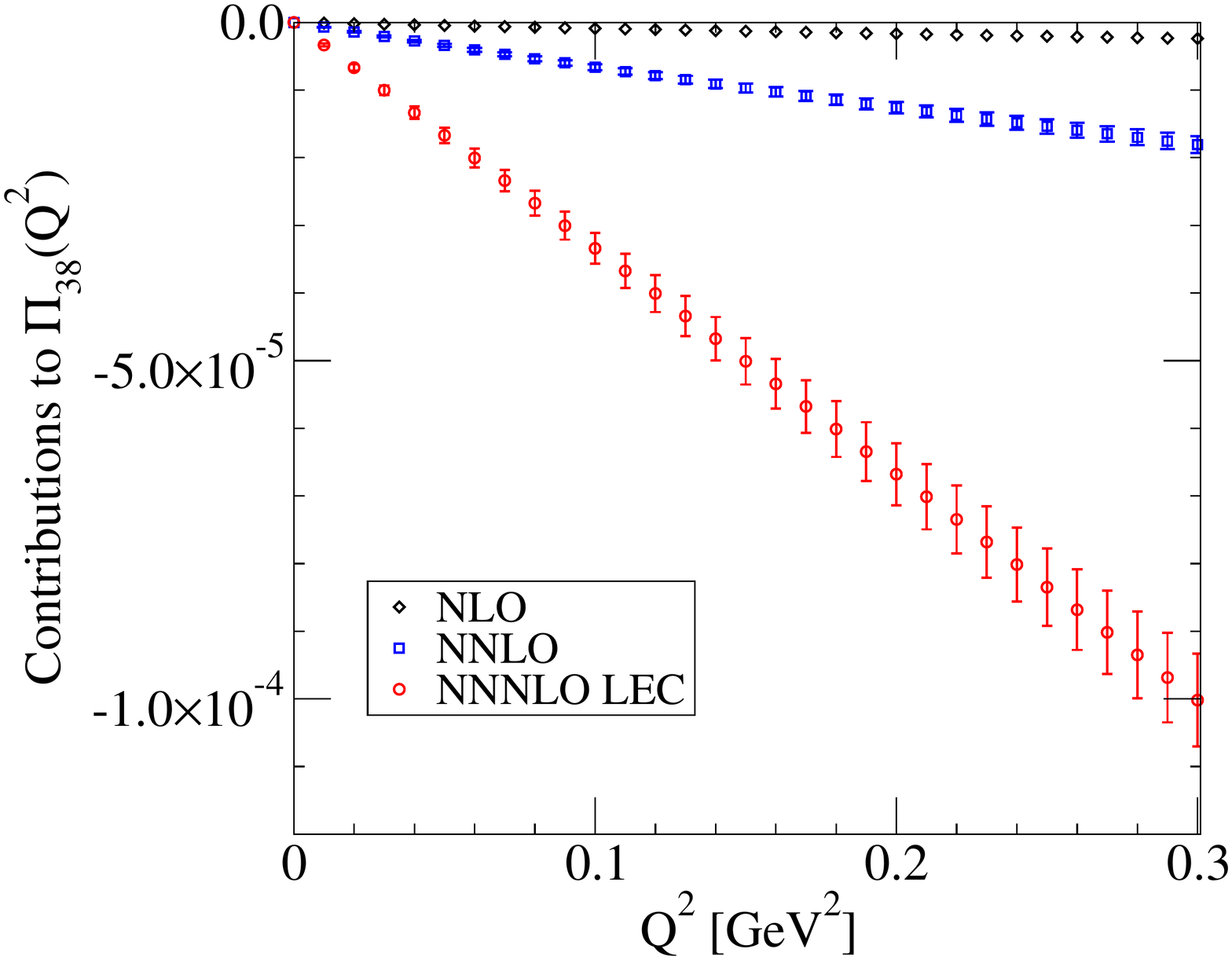}
{\caption{\label{nlonnlonnnlocontsvsQ2}} The NLO, NNLO and NNNLO LEC
contributions to $\hat{\Pi}^{38}(Q^2)$. The errors on the NNLO and NNNLO 
LEC points are those induced by the uncertainties on the input value for 
$L_9^r$ and the contribution to the error on $\delta C_{93}^{(1)}$ 
quoted in Eq.~(\ref{updateddelc931}), respectively.}
\end{figure}
\end{center}

Figure~\ref{nlonnlonnnlocontsvsQ2} shows the $Q^2$ dependence of the
NLO, NNLO and ``NNNLO LEC'' contributions to $\hat{\Pi}^{38}(Q^2)$, where
``NNNLO LEC'' denotes the tree-level contribution proportional to
$\delta C_{93}^{(1)}$. It is clear that the NNNLO LEC contribution is 
numerically dominant, and that, although the loop functions which
determine the NLO and NNLO contributions are not strictly linear in $Q^2$,
they are, numerically, very close to being so, in the region of interest 
to us. The errors on the NNLO and NNNLO LEC contributions are those
associated with the uncertainty on the input for $L_9^r$, and that
on the leading-order-in-FB result, Eq.~(\ref{updateddelc931}),
for $\delta C_{93}^{(1)}$. 

Adding to the NNNLO LEC result, (\ref{nnnlolecamusibresult}), the NLO and 
NNLO contributions (\ref{nlosibcont}) and (\ref{nnlosibcont}), we obtain 
our final estimate for $a_\mu^{SIB}$,
\begin{equation}
a_\mu^{SIB}=3.32(4)(19)(33)(81)\times 10^{-10}\, ,
\label{finalamusib}\end{equation}
where the first error is that induced on the NNLO contribution by the 
uncertainty on the input for $L_9^r$, the second is that associated with 
the error on the FB IMFESR estimate, Eq.~(\ref{updateddelc931}), for 
$\delta C_{93}^{(1)}$, the third is our $10\%$ estimate for the uncertainty 
produced by the combination of truncating the integral for $a_\mu^{SIB}$ at 
$Q^2_{max}=0.25\ GeV^2$ and neglecting contributions beyond NNLO to the 
curvature of $\hat{\Pi}^{SIB}(Q^2)$, and the fourth is that induced by 
our $\sim 30\%$ estimate for the uncertainty in $\delta C_{93}^{(1)}$ 
induced by possible higher-order FB contributions to the slope of 
$\hat{\Pi}_{ud-us;V}(Q^2)$ at $Q^2=0$ obtained from the updated
version of the FB IMFESR analysis of Ref.~\cite{Golterman:2017ljr}.

\begin{center}
\begin{figure}[h]
\includegraphics[width=.8\textwidth,angle=0]
{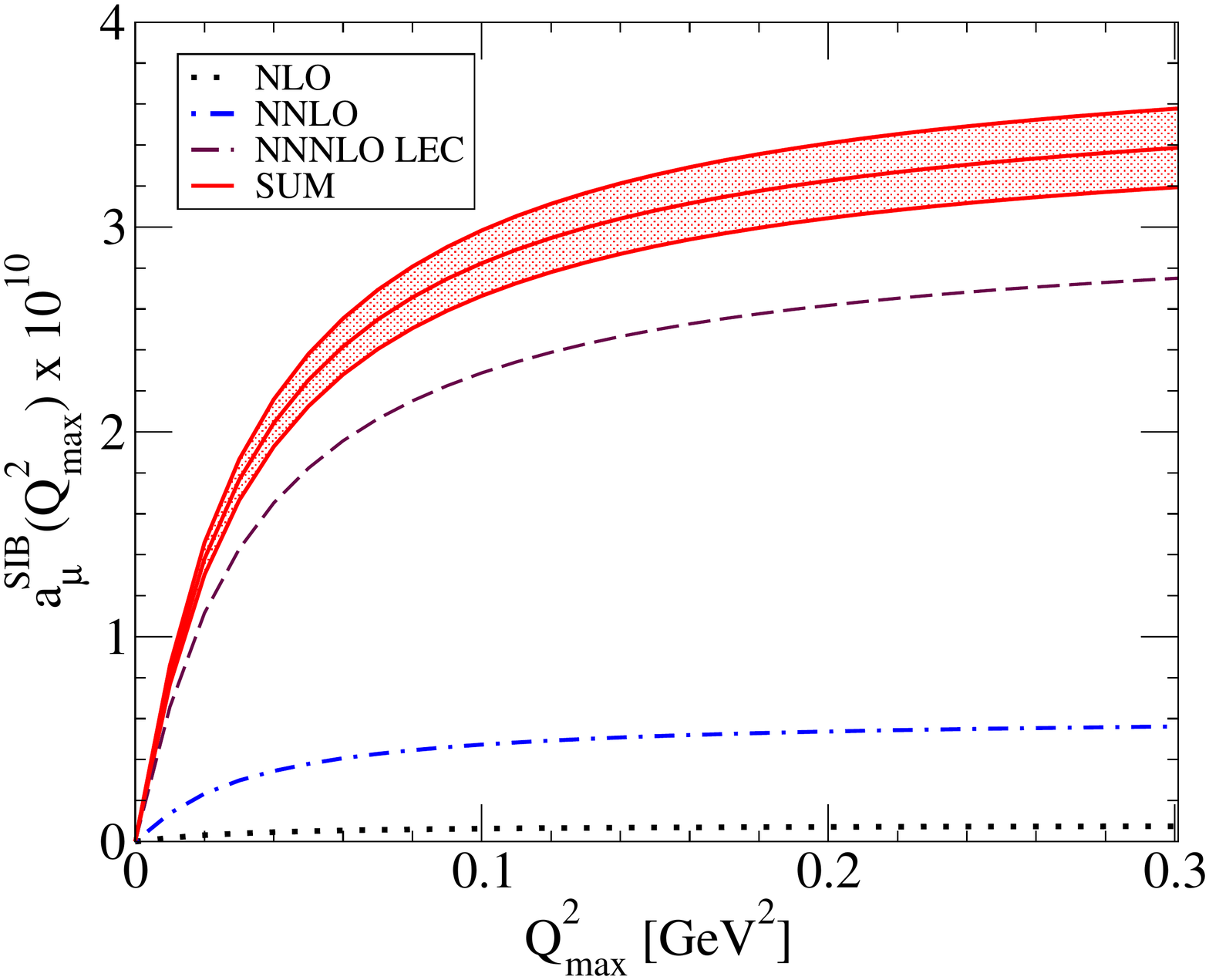}
{\caption{\label{amusigaccumulation} The accumulation of the
NLO, NNLO and NNNLO LEC contributions to $a_\mu^{SIB}$ as a
function of the upper integration limit, $Q^2_{max}$. The errors
on the NNLO and NNNLO LEC contributions have been suppressed.
The shaded band shows the error on the sum of the NLO, NNLO and NNNLO LEC 
contributions obtained by summing the NNLO and NNNLO LEC errors from
Fig.~\ref{nlonnlonnnlocontsvsQ2} in quadrature.}}
\end{figure}
\end{center}

The NLO, NNLO and NNNLO LEC contributions to $a_\mu^{SIB}[Q^2_{max}]$,
together with the NLO+NNLO+NNNLO LEC total, are shown as a function of 
$Q^2_{max}$ in Figure~\ref{amusigaccumulation}. The error band on the total
shows the quadrature sum of the LEC-uncertainty-induced NNLO and NNNLO LEC 
errors plotted in Fig.~\ref{nlonnlonnnlocontsvsQ2}.

\section{\label{conclusions}Summary and conclusions}
We have obtained a continuum, ChPT-based estimate of the SIB contribution, 
$a_\mu^{SIB}$, to $a_\mu^{LO,HVP}$, the leading-order, 
hadronic-vacuum-polarization contribution to the anomalous magnetic 
moment of the muon. As shown in Figs.~\ref{nlonnlonnnlocontsvsQ2} and 
\ref{amusigaccumulation}, the NLO contribution to this result is very small, 
presumably as a consequence of the cancellation at this order between 
disconnected and connected contributions from $\pi\pi$ intermediate states. 
The NNLO contribution, though significantly larger, is also sub-dominant, 
a result not unexpected given the absence of terms encoding resonance-region 
contributions from the NNLO representation. Resonance-region contributions
first appear in the chiral expansion of $a_\mu^{SIB}$ at NNNLO, encoded in
the NNNLO LEC $\delta C_{93}^{(1)}$. Our full estimate, (\ref{finalamusib}), 
for $a_\mu^{SIB}$ is thus, as expected, dominated by the NNNLO contribution 
proportional to $\delta C_{93}^{(1)}$. Fortunately, an estimate for this 
LEC can be obtained from a FB IMFESR analysis of experimental hadronic 
$\tau$ decay distributions, and we have updated the original version of
this analysis, reported in Ref.~\cite{Golterman:2017ljr}, to take into
account subsequent, numerically relevant changes to the normalization of 
the dominant $K\pi$ contribution to the strange experimental distribution. 
The resulting NNNLO LEC contribution to $a_\mu^{SIB}$ is similar in size
to the results of phenomenological estimates for the contribution from the 
$\rho$-$\omega$ interference region based on model-dependent fits to 
experimental interference-region $e^+ e^-\rightarrow \pi^+\pi^-$ 
cross sections, confirming the importance of contributions from the
$\rho$-$\omega$ region. The ChPT analysis has the advantage, over such
phenomenological estimates of the contribution from this one, 
narrow region only, of including also contributions from the lower-$Q^2$ 
region, evaluated in the model-independent chiral framework, as well as
those from regions of the spectrum above $s\simeq m_\omega^2$ where the 
absence of experimentally observable IB interference effects makes 
analogous phenomenological estimates impossible. 
 
The dominance of the result in Eq.~(\ref{finalamusib}) by the 
NNNLO LEC term in the chiral representation of $\hat{\Pi}^{38}(Q^2)$ 
and hence by contributions from higher-energy (short-distance) resonance 
degrees of freedom confirms the expectation that, once connected and
disconnected contributions have been summed, FV effects in lattice 
determinations of $a_\mu^{SIB}$ will be small, relative to $a_\mu^{SIB}$,
and hence can be neglected on the scale of the current precision goal 
for the determination of $a_\mu^{LO,HVP}$. The situation for the relative 
size of FV effects should, in fact, be similar to that of the $I=1$ 
contribution, $a_\mu^{33}$, where the contribution proportional to the
NNLO LEC $C_{93}$ which encodes the higher-energy $\rho$ degree of
freedom also dominates the chiral representation. The only difference
 between the two cases is a practical one: while few-to-several percent 
FV corrections to the large $a_\mu^{33}$ contribution are far from 
numerically negligible on the scale of the current precision target, 
analogous few-to-several percent FV corrections to the much (more 
than two orders of magnitude) smaller SIB contribution are entirely 
negligible on that same precision target scale.

Combining the errors from Eq.~(\ref{finalamusib}) in quadrature, we find
for our final result
\begin{equation}
a_\mu^{SIB}\, =\, 3.32(89)\times 10^{-10}\, .
\label{finalquadsumerramusib}\end{equation}
The central value is larger than that of the BMW lattice result, 
\begin{equation}
\left[a_\mu^{SIB}\right]_{BMW}\, =\, 1.93(83)(87)\times 10^{-10}
\, =\, 1.93(1.20)\times 10^{-10}
\label{bmwamusib}\end{equation}
obtained by summing the connected and disconnected contributions 
reported in Ref.~\cite{Borsanyi:2020mff}, but compatible with it 
within errors.{\footnote{The statistical and systematic errors, 
$0.83\times 10^{-10}$ and $0.87\times 10^{-10}$, on the BMW result
are the quadrature sums of the corresponding statistical/systematic 
errors on the connected and disconnected contributions. We thank 
Laurent Lellouch for clarification on how these errors should
be combined.}}

We close by noting that, given the dominance of the result by the 
contribution proportional to the NNNLO LEC $\delta C_{93}^{(1)}$, and
the leading linear-in-$Q^2$ behavior of this contribution, it would be 
of interest were future lattice studies to quote results for the slope 
of $\hat{\Pi}^{SIB}(Q^2)$ with respect to $Q^2$ at $Q^2=0$, a result 
otainable from the $t^4$ time moment of the two-point function at 
zero spatial momentum~\cite{Chakraborty:2014mwa}.

\begin{acknowledgments}
The work of C.L.J., R.L. and K.M. is supported by the Natural
Sciences and Engineering Research Council of Canada
\end{acknowledgments}

%%%%%%%%%%%%%%%%%%%%%%%%%%%%%%%%%%%%%%%%%%%%%%%%%%%%%%%%%%%%%%%%%%%%%%%%%

\vfill\eject
\end{document}